\input{aipcheck}
\documentclass[
    ,final            
  ]
  {aipproc}

\layoutstyle{6x9}

\begin{document}

\title{On the relaxation dynamics of glass-forming systems: Insights from
computer simulations}


\author{Pinaki Chaudhuri}{
  address={Laboratoire des Collo{\"\i}des, Verres
et Nanomat{\'e}riaux, UMR 5587, Universit{\'e} Montpellier 2 and CNRS,
34095 Montpellier, France}
}

\author{Ludovic Berthier}{
  address={Laboratoire des Collo{\"\i}des, Verres
et Nanomat{\'e}riaux, UMR 5587, Universit{\'e} Montpellier 2 and CNRS,
34095 Montpellier, France}
}

\author{Srikanth Sastry,}{
  address={Jawaharlal Nehru Centre for Advanced Scientific Research ,
Jakkur Campus, Bangalore 560064, India}
}

\author{Walter Kob}{
  address={Laboratoire des Collo{\"\i}des, Verres
et Nanomat{\'e}riaux, UMR 5587, Universit{\'e} Montpellier 2 and CNRS,
34095 Montpellier, France}
}

\begin{abstract}
We discuss the relaxation dynamics of a simple lattice gas model for
glass-forming systems and show that with increasing density of particles
this dynamics slows down very quickly. By monitoring the trajectory
of tagged particles we find that their motion is very heterogeneous
in space and time, leading to regions in space in which there is a
fast dynamics and others in which it is slow. We determine how the
geometric properties of these quickly relaxing regions depend on density
and time. Motivated by this heterogeneous hopping dynamics, we use a
simple model, a variant of a continuous time random walk, to characterize
the relaxation dynamics. In particular we find from this model that for
large displacements the self part of the van Hove function shows an
exponential tail, in agreement with recent findings from experiments
and simulations of glass-forming systems.
\end{abstract}

\maketitle

\section{Introduction}

\begin{figure}[b]
{\includegraphics[scale=1.0]{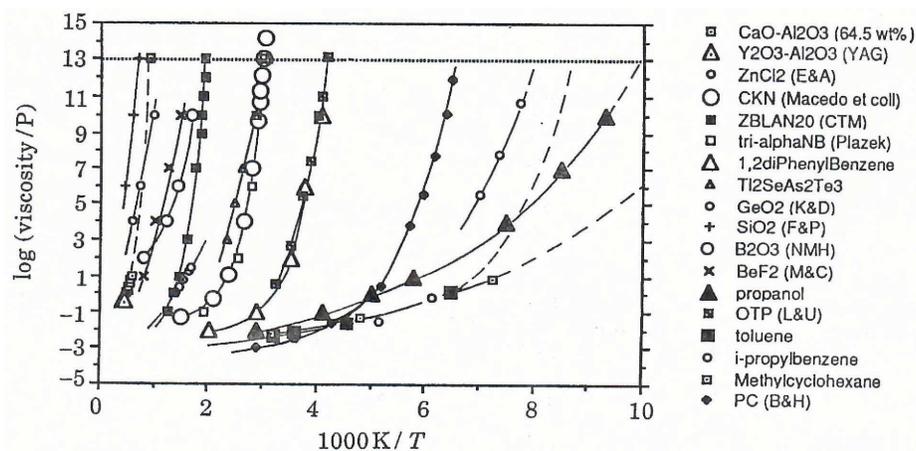}}
\caption{Arrhenius plot of the temperature dependence of the
viscosity of various glass-forming liquids. After Ref~\cite{angell94}.
}
\label{fig1_eta}
\end{figure}

Matter occurs in three different phases: Gas, crystal, and liquid. In a
gas the particles are separated by distances that are large with respect
to the size of the particles (or more precisely with respect to the
typical length-scale for the interaction between the particles). Therefore
density is a small parameter which allows an
accurate theoretical description of the structure and dynamics of these
systems. In a crystal the long range structural order permits to use
periodic functions in space and time to describe the lattice vibrations
(phonons) of the system and thus we have a quite accurate understanding of
the structural, dynamical and mechanical properties of crystals. Things are
more difficult in liquids: Their density is comparable to the ones of crystals
and hence cannot be used as small parameter. Their structure is neither
completely disordered as in a gas nor ordered as in a crystal. Therefore
the investigation of structural and dynamical properties of liquids
remains an active field of research, despite the knowledge
that we have on these systems~\cite{barrat03,hansen07}. One of the
most puzzling behavior of liquids is the temperature dependence of
their viscosity $\eta$. In Fig.~\ref{fig1_eta} we show $\log(\eta)$
for a wide variety of liquids as a function of $1/T$. Note that in this
representation an Arrhenius dependence, $\eta(T) \propto \exp(E/k_BT)$
(with $E$ the activation energy and $k_B$ the Boltzmann constant),
would just be a straight line. Several important conclusions can be
drawn from this graph: 1) The $T-$dependence of the viscosity is
extremely steep in that a change of $T$ by a factor of 2-3 can lead
to an increase of $\eta$ by 12-15 decades. 2) This $T-$dependence is
non-Arrhenius in that most of the curves are bent upwards, i.e. $\eta$
shows a super-Arrhenius behavior. 3) The curves show a completely smooth
$T-$dependence. In particular one does not see any discontinuity around
the melting temperature of the material. (We note that for most of these
liquids the melting occurs at temperatures at which $\eta(T)$ is on the
order of $10^0-10^4$Pa~s.) Furthermore we mention that for many liquids
there is no experimental data for the viscosity for temperatures that
are significantly below the melting temperature, since these systems
crystallize if they are supercooled. However, as the figure shows, there
is indeed a large class of liquids, the so-called ``glass-formers'' for
which the crystallization process is so slow that $\eta$ can be measured.

Finally we note that the $T-$dependence of $\eta$ seems to extrapolate
smoothly to even lower temperatures. In view of the experimental fact
that the structural relaxation time of a liquid, as measured, e.g., by
means of the intermediate scattering function~\cite{hansen07}, shows a
$T-$dependence that is very similar to the one of $\eta$, one thus can
conclude that there will be a temperature below which the relaxation time
becomes so large that the sample will no longer flow on human time scales,
i.e. it has become a {\it glass}.  Despite large efforts to identify the
mechanism(s?) that are responsible for the dramatic slowing down of the
dynamics of these liquids, and hence of the resulting glass-transition,
there is presently no theory that is able to give a reliable description
of this dynamics~\cite{gotze92,gotze99,debenedetti_01,das04,binder05}.

The strong increase of the viscosity (or the relaxation times)
is not the only puzzling feature of glass-forming liquids. If one
investigates these systems on a more microscopic scale, e.g. by using
light or neutron scattering techniques, one finds that typical time
correlation functions, $\phi(t)$, are not given by a Debye-law,
i.e. an exponential decay in time, but are instead stretched,
i.e. $\phi(t) \propto \exp(-(t/\tau)^\beta)$, with a stretching exponent
$\beta<1.0$~\cite{binder05,ediger96}. One possibility to explain this
stretching phenomenon is the presence of the so-called ``dynamical
heterogeneities''~\cite{richert94,sillescu99,ediger00,richert02} in the
glass-formers.  What do we mean by this? Since the system is disordered,
each particle will have a somewhat different environment. Hence it
can be expected that each particle has a somewhat different relaxation
dynamics. Some of them relax faster, some of them slower. This is the
qualitative meaning of dynamical heterogeneity. (Of course on the long
run the environment of a particle will change and hence a particle that
relaxes quickly will become slow and vice versa, i.e. at very large times
all particles will have the same statistical properties.) Since in most
experiments one is only able to measure the {\it average} relaxation
dynamics of the particles, one will average over (momentarily) fast
and slow particles and hence the average time correlation function
is not a single exponential, but a superposition of exponentials
with different relaxation times, and such an average can usually be
described well by a stretched exponential. (The exact form of $\phi(t)$
will depend on its exact nature and on the distribution of fast and slow
particles.) Experiments and simulations have shown that these dynamical
heterogeneities do indeed exist in that there are regions in the sample
in which the particles move quickly and others in which they move
slowly~\cite{ediger00,richert02,kob97,kegel00,weeks00,berthier04,appignanesi06}.

Despite this success to rationalize, at least qualitatively,
the stretching, it is presently not very clear to what extent these
dynamical heterogeneities are really responsible for understanding the
relaxation dynamics of glass-forming liquids~\cite{appignanesi06}. 
Similarly, the nature of the motion of the
particles in the mobile/immobile regions is not well understood. 
In order to find answers to
these questions we have done computer simulations of various glass-forming
systems and in the following we will discuss some of the results that
have helped to clarify matters at least to some extent. In the first part
of this brief review we will discuss the results obtained for a simple lattice
gas in which we were able to characterize the nature of the dynamical
heterogeneities in great detail. In the second part we will show that the
presence of these dynamical heterogeneities have a surprising consequence
for certain time correlation functions and will present a simple model,
the continuous time random walk, that is able to capture some essential
features of these correlation functions.

\section{Model and Details of the Simulations}

As discussed in the previous section, the relaxation dynamics of
glass-forming systems spans at least 12-15 decades in time. Despite the
availability of supercomputers, it is currently impossible to do computer
simulations of, say, a system of O($10^3$) particles, which extend
over that many decades in time. Therefore one has to restrict oneself
to a smaller time window and in addition to use interaction potentials
that are as simple as possible in order to maximize numerical efficiency.
One class of models that has been found to be very
useful for the study of glassy dynamics are the so-called lattice gases
(or kinetic Ising models) in which the particles (or defects) move on a
regular lattice (see Ref.~\cite{ritort03} for a review on these systems).

In the present work we will use one particular case of such a
lattice gas, the so-called KA model~\cite{kob93} which is defined
as follows: $N$ particles populate a cubic lattice of size $L^3$
with the constraint that a lattice site can be occupied by only one
particle. All possible configurations have the same energy and thus the
same Boltzmann weight. This property implies that the system does not
require equilibration since every randomly generated configuration has
the same statistical probability, a feature which is most useful for
computer simulations. The imposed stochastic dynamics consists of the
following process: A randomly selected particle can move to any one
of the neighboring empty lattice site provided it has $m$ or fewer
occupied nearest neighbor sites and that the target empty site has
$m + 1$ or fewer occupied nearest neighbor sites. A choice of $m=3$
results in a dynamics that, at high densities $\rho=N/L^3$, shares
many properties of real glass-forming systems, such as stretching
of the time correlation functions, dynamical heterogeneities,
etc.~\cite{kob93,sellitto02,franz02,berthier03,toninelli04,toninelli05,marinari05}.

For efficient sampling of the configuration space at high densities,
we have carried out event-driven Monte Carlo \cite{kob93,bortz75}
simulations of the model. In the following we will report the results
using the lattice spacing as a unit of length and one Monte Carlo sweep
of the system as the unit of time.  Using periodic boundary conditions,
we have investigated system sizes $L=20$, 30, and 50, which avoid effects
due to finite size~\cite{kob93}, with densities spanning from $\rho=0.65$
to $\rho=0.89$.

\section{Results}
In this section we will discuss the relaxation dynamics of the
KA-model. Firstly we will show that this dynamics is indeed
quite similar to the one found in more standard (off-lattice)
glass-formers. Subsequently we will discuss the dynamical heterogeneities
and show how they are related to the relaxation dynamics.

As mentioned above, in this model all possible configurations of
particles have the same statistical weight. Therefore the static
structure is trivial and shows no relevant $\rho-$dependence, i.e. in the
thermodynamic limit structural quantities like the radial distribution
function etc. will depend only on $\rho^{1/3}$. Therefore we can
concentrate in the following on the $\rho-$dependence of the dynamical
quantities. One important quantity is the so-called intermediate
scattering function $F_s(q,t)$ which is defined as~\cite{hansen07}

\begin{equation}
F_s(q,t) = \frac{1}{N} \sum_{j=1}^N \langle \exp(i \vec{q} \cdot 
(\vec{r}_j(t)-\vec{r}_j(0))) \rangle \quad .
\label{eg1}
\end{equation}

\noindent
Here $\vec{r}_j(t)$ is the position of particle $j$ at time $t$,
$\vec{q}$ is a wave-vector, and $\langle . \rangle$ is the average over
the configurations. (Note that due to rotational symmetry, the left hand
side depends only on $q=|\vec{q}|$.)

\begin{figure}[thb]
{\includegraphics[scale=0.35]{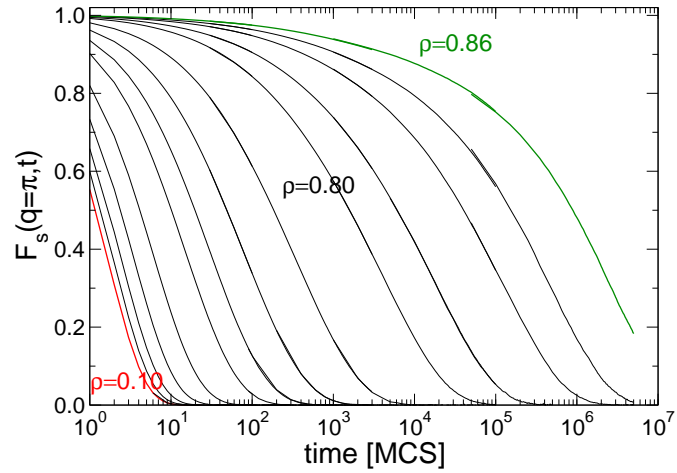}}
\caption{Time dependence of the incoherent intermediate scattering
function for different densities $\rho$. From left to right: $\rho=0.1$,
0.2, 0.3, 0.4, 0.5, 0.6, 0.65, 0.7, 0.75, 0.8, 0.82, 0.84, 0.85, 0.86.
}
\label{fig2_fs}
\end{figure}

In Fig.~\ref{fig2_fs} we show the time dependence of $F_s(q,t)$ for
different densities $\rho$. The wave vector is $q=\pi$, i.e. one measures
the displacement of the particles over a distance which is on the order
of one lattice spacing. (Other wave-vectors show a similar $t$ and
$\rho-$dependence.)  From this figure we recognize that at low densities
the relaxation dynamics is very fast in that the time correlation function
decays to zero within 10 Monte Carlo steps (MCS). With increasing $\rho$
the dynamics slows down considerably and at $\rho=0.86$ it takes more than
$10^7$ MCS before $F_s(q,t)$ has decayed to zero. Furthermore we note
that at the highest densities the shape of the correlator is no longer
an exponential, but instead can be fitted well by a stretched exponential
with a stretching exponent $\beta$ around 0.63~\cite{kob93}. Accompanied
with this slowing down is a quick decrease of the diffusion constant
for the particles (which can, e.g., be calculated from the mean squared
displacement and the Einstein relation)~\cite{kob93}. Hence these results
show that, despite its simplicity, this lattice gas model shows some of
the typical features of glass-forming systems.

\begin{figure}[b]
{\includegraphics[scale=0.40,trim = 10mm 00mm 20mm 40mm, clip]{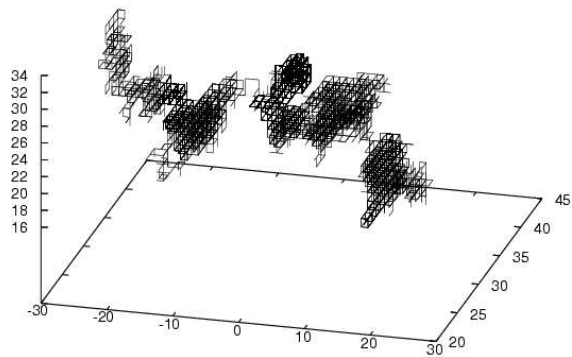}}
\caption{Typical trajectory of a tagged particle that is mobile. Total length of the
time is $10^7$ MCS and the density is $\rho=0.87$.
}
\label{fig3_traject}
\end{figure}

\begin{figure}[t]
{\includegraphics[scale=0.35,angle=270]{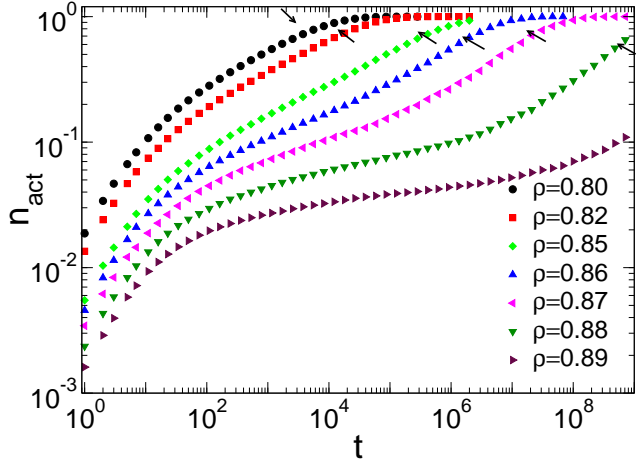}}
\caption{
Time dependence of the concentration of active sites for different
densities.  The arrows mark the $\alpha-$relaxation time $\tau_\alpha$.
}
\label{fig4_nact}
\end{figure}

How do the particles explore space? In order to find an answer to this
question we show in Fig.~\ref{fig3_traject} the trajectory of one tagged
particle. The time span is $10^7$ MCS and the density is $\rho=0.87$. As
we can recognize from the figure the trajectory consists of relatively
compact blobs, having a typical diameter of 6, that are connected by
means of narrow pathways. Note that this structure is very different
from the one expected for a simple random walk trajectory on
a lattice for which, in three dimensions, the walk does not form such
compact regions. Thus the particle explores a first blob, finds one of
the pathways that leads to a neighboring blob, explores that one, moves
on, etc. Thus we can conclude that there is a microscopic mechanism
in the system that makes the relaxation dynamics very heterogeneous,
i.e. non-uniform in space, and below we will discuss this point in
more detail.

The fact that the tagged particle is able to explore the observed
blobs implies that the other particles that are in these blobs have
to move as well. Therefore this result gives us a first indication
that the dynamics of this system is not only heterogeneous, but also
cooperative, qualitatively similar to results found in other glass-forming
systems~\cite{sillescu99,ediger00,richert02,appignanesi06,donati98}.

A further interesting observation is that if one considers the trajectory
of a second tagged particle that, at $t=0$ is within one of the blobs
explored by the first particle, one finds that
(generally) the former trajectory has a very strong overlap with the
second one~\cite{chaudhuri08}. Thus it seems that there are lattice
sites in the system that have a high throughput of particles
whereas there are other sites in which there is no action, i.e. at these
latter sites one finds for a very long time always the same particle
(or, in rare exceptions, the same vacancy). Thus we can conclude that the
relaxation dynamics of the system must be related to the presence of
these sites with high traffic of particles.  In order to investigate
this point further we define a site as ``active'' if within the time
interval $[0,t]$ it has been visited by more than one particle
or vacancy. We emphasize that the presence of such active sites must
be intimately related to the initial configuration of the particles,
since, of course, on very long time scales all lattice sites must have
exactly the same statistical properties (e.g. have had the same number
of particles that passed through the site, etc.).

\begin{figure}[thb]
{\includegraphics[scale=0.90,angle=270]{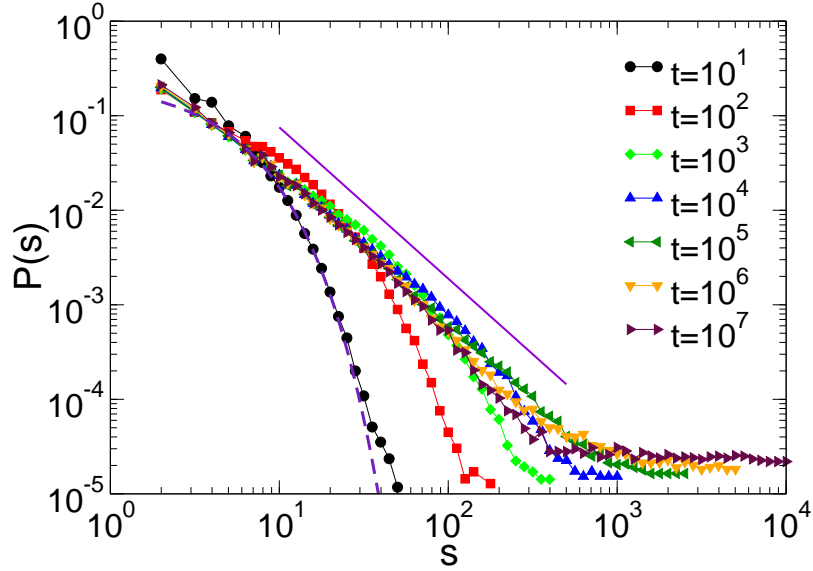}}
\caption{
Distribution of cluster size, $P(s,t)$, where $s$ is the size of
cluster, for different times at density $\rho=0.88$. Also shown (dashed
line) is an exponential fit to $P(s,t)$ at $t=10$ and (solid line) and
the function ${s^{-1.6}}$ to demonstrate the emergence of power-law
behavior of $P(s,t)$ at intermediate times.
}
\label{fig5_cluster}
\end{figure}

In Fig.~\ref{fig4_nact} we show how the concentration of active sites
$n_{act}(t)$, i.e. the total number of active sites divided by $L^3$, depends on
time and density. For short times $n_{act}(t)$ increases rapidly and we
find $n_{act}(t)\approx \alpha(\rho) (1-\exp(-t/\theta))$ with $\theta\approx5$, independent
of $\rho$, although the prefactor $\alpha(\rho)$ of this $t-$dependence decreases with
$\rho$.  This regime corresponds to the initial (fast) growth of the
blobs, whose concentration decreases with $\rho$.  Subsequently the shape
of $n_{act}(t)$ depends strongly on $\rho$. For low densities the curves
approach rapidly $n_{act}=1$, i.e. all the sites have seen more than
one particle/vacancy, and thus the system has completely relaxed. This
approach can be described well by a stretched exponential tail, with
a stretching exponent of around $0.6$, a functional form that is found
also for the higher values of $\rho$. For high densities one finds a third
regime in that $n_{act}(t)$ shows at intermediate times a very slow growth.
This intermediate regime corresponds to the slow growth and coalescence
of the blobs that will with time form the percolating structure seen
in Fig.~\ref{fig3_traject}. Note that the growth rate of these blobs
decreases rapidly with increasing $\rho$ since the number of vacancies
that are neighbors of the blob and which help it to grow decreases with
increasing density. Also included in the figure is, for each density,
the $\alpha-$relaxation time $\tau_\alpha$ (small arrows) which is
defined as the time it takes for the intermediate scattering function
to decay to $e^{-1}$. As we can see, $\tau_\alpha$ corresponds to a
time scale in which $n_{act}(t)$ has already entered the final stretched
exponential decay, i.e. most of the sites have become active. Hence, the
very slow growth at intermediate times seen at high $\rho$ is not the
$\alpha-$relaxation of the whole system. However, it plays a 
significant role in the observed slowing down of dynamics prior to final 
relaxation.

\begin{figure}[thb]
{\includegraphics[scale=0.90,angle=270]{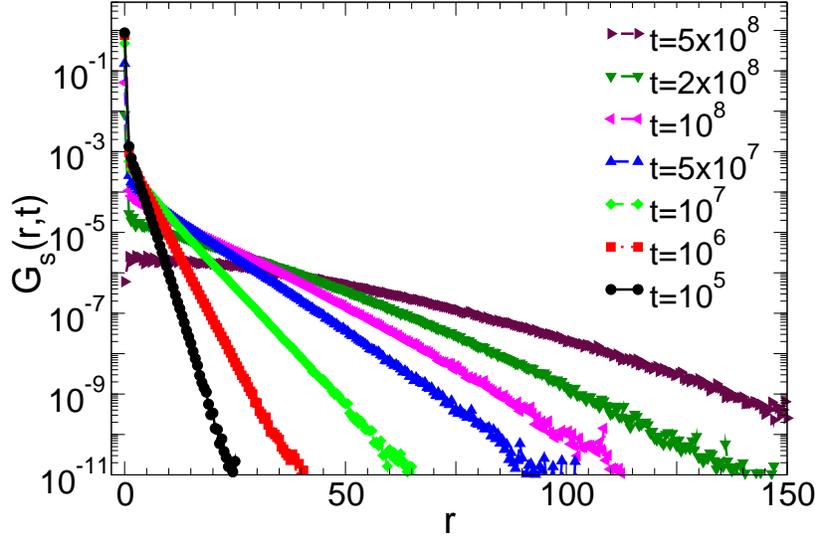}}
\caption{
$G_s(r, t)$ for different times at a density of $\rho = 0.87$.
The $\alpha-$relaxation time at this density is $\tau_\alpha \approx 1.8
\times 10^7$.
}
\label{fig6_vanhove}
\end{figure}

In order to characterize the geometric structure formed by the active
sites we have determined $P(s,t)$, the distribution of the size $s$ of the
clusters formed by these active sites at time $t$. (We define two active
sites to belong to the same cluster if they can be connected to each
other via a sequence of active sites that are nearest neighbors.) This
distribution is shown in Fig.~\ref{fig5_cluster} for the (high) density
$\rho=0.88$. For short times $P(s,t)$ can be approximated well by an
exponential distribution (dashed line) since at $t=0$ the active sites
will be distributed randomly. With increasing $t$ the geometric structure
of the clusters starts to evolve towards a percolating backbone, i.e. the
concentration of large clusters will grow. For intermediate times the
distribution for small and intermediate $s$ can be approximated well by a
power-law, $P(s,t) \propto s^{-\nu}$, with a fractal exponent $\nu \approx
1.6$, which shows that the percolating structure has fractal nature. For
very large $s$ the distribution becomes flat, which is most likely a
finite size effect related to the periodic boundary conditions. Note that
the times shown in Fig.~\ref{fig5_cluster} are all significantly shorter
than the $\alpha-$relaxation time $\tau_\alpha$ which, for this $\rho$,
is around $4.4\times10^8$. For times of the order of $\tau_\alpha$ and
beyond (not shown), the distribution $P(s,t)$ shows a peak at $s\approx
L^3$, i.e. most of the sites in the system have become active.

Figures \ref{fig3_traject} and \ref{fig5_cluster} show that the relaxation
dynamics of the system is very heterogeneous in space and time. Thus
it can be expected that there are certain particles that are relatively
fast and others that are relatively slow. That this is indeed the case
can be demonstrated by measuring the self part of the van Hove function
$G_s(r,t)$ which is defined as~\cite{hansen07}

\begin{equation}
G_s (r,t)= \frac{1}{N} \sum_{j=1}^N \langle
\delta ( r - |{\bf r}_j(t) - {\bf r}_j(0)|) \rangle \quad .
\label{eq2}
\end{equation}

\noindent
Thus $G_s(r,t)$ gives the probability that a particle makes within the
time interval $t$ a displacement of size $r$.  In Fig.~\ref{fig6_vanhove}
we show this distribution function for $\rho=0.87$. From this figure
we recognize that for short and intermediate times the distribution
has a strong peak at $r=0$, i.e. most of the particles have not
moved at all. Even for times as large as $t=10^5$ the fraction of
particles that has moved a distance $r\geq 1$ is very small, only around $1\%$.
If one looks at displacements $r\ge 1$ one finds that there
are, e.g. for $t=10^7$, even a few particles that have moved a distance as large
as 50! Also noticeable is the fact that at intermediate times $G_s(r,t)$
is not given by a Gaussian distribution, as one would expect for a
tracer particle that undergoes Brownian diffusion, but instead by an
exponential distribution. Only for times that are much longer than the
$\alpha-$relaxation time, which at this density is around $1.7 \times
10^7$, we find that $G_s(r,t)$ becomes a Gaussian, i.e. one has to go
to very long time scales in order to observe a dynamics that is really
random. From this figure we thus can conclude that those particles that
can move already on time scales on the order of $\tau_\alpha$, do not seem
to do this like a randomly diffusing particle. Instead the presence of
the blobs, and at later time the percolating structure of the backbone,
makes that the relaxation dynamics of these mobile particles obeys a
different statistics. In the following section we will discuss this motion
in more detail and also present a simple model that is able to describe it
quantitatively. For the moment we thus just conclude that the relaxation
of this simple glass-former happens in a hierarchical manner: First there
is a fast local relaxation within blobs. Subsequently these blobs grow
with a rate that depends strongly on density. The slow coalescence of
these blobs lead to the formation of a percolating network. This backbone
fattens until it finally occupies the whole lattice, which corresponds
to the $\alpha-$ relaxation of the system. Hence we can conclude that
the relaxation dynamics of this simple system does indeed show spacial
and temporal heterogeneities, qualitatively similar to the ones that are
(indirectly) observed in real glass-formers.

\section{A Continuous Time Random Walk Description of the Dynamics}

In this section we will present a simple model that is able to give a
quantitative description of the relaxation dynamics, as described by
the self part of the van Hove function, of glass-forming systems.

\begin{figure}[thb]
{\includegraphics[scale=0.9]{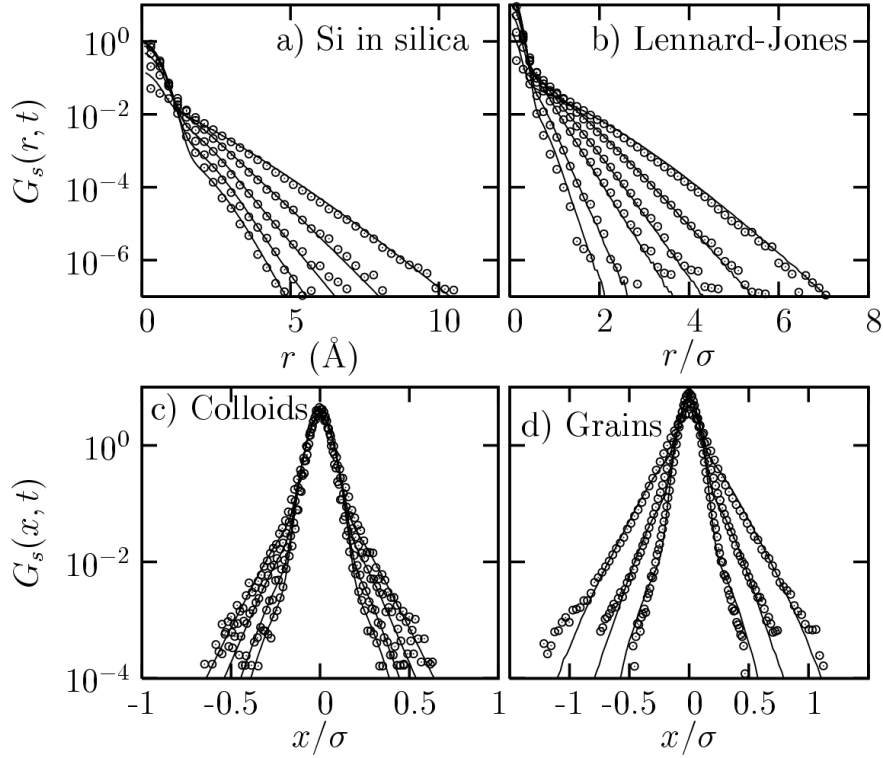}}
\caption{
Self part of the van Hove function for different times and different
systems. a) Silicon in a silica system. $T=3000$~K and $t \in
[27,1650]$ps. b) Lennard-Jones particles at $T=0.435$ and $t \in
[7.5\times 10^4, 4.1 \times 10^7]$. c) Colloidal hard spheres at a
packing fraction $\phi=0.517$ and $t \in [90,1008]$s. d) A granular
system at $\phi=0.84$ and $t \in [10,1000]$ cycles. a) and b) show
the distribution of $|\vec{r}(t)-\vec{r}(0)|$, and c) and d) show the
distribution of $x(t)-x(0)$.
}
\label{fig7_vanhoveall}
\end{figure}

In order to have a model that is not only applicable to the lattice gas
system discussed so far, we will consider a general glass-former, i.e. an
off-lattice system. In a first step we thus have to understand the shape
of the self part of the van Hove function for off-lattice systems. This
function is shown in Fig.~\ref{fig7_vanhoveall} for four different
glass-formers. The data for silica, panel a), comes from a molecular
dynamics simulation~\cite{sio2} 
with the so-called BKS-potential~\cite{van_beest90},
an interaction potential that has been shown to reproduce many structural
and dynamical properties of real silica~\cite{vollmayr96,horbach99}. Panel
b) shows data for a binary Lennard-Jones system~\cite{LJ}, a model which in the
past has been investigated intensively and has been found to share
many properties of real simple glass-formers~\cite{kob94,kob99}. The
colloidal data shown in panel c) have been obtained by moni\-toring a
colloidal suspension of hard sphere particles by means of a confocal
microscope and which thus allows to track directly the trajectories of
the particles~\cite{weeks00}.  Finally panel d) shows the data from
a two-dimensional granular system in which the particles are shuffled
horizontally via oscillating walls~\cite{marty05}. From this figure we
can conclude that for all these glass-forming systems the distribution
$G_s(r,t)$ shows a Gaussian central part and an exponential tail at
larger distances. The only difference between these off-lattice systems
and the lattice gas discussed above is thus that the latter system shows a
$\delta-$peak at $r=0$ whereas in the former systems this peak is replaced
by a Gaussian distribution. Similar results have also been found in other
glass-forming systems~\cite{stariolo06,chaudhuri08b}. This finding is
of course very reasonable, since in an off-lattice system the particles
will, for short times, just oscillate around their (temporal) equilibrium
position since they are trapped by their nearest neighbors. To a good
approximation the local potential generated by the neighbors is given by
a quadratic form, i.e. the oscillations of the particles is harmonic,
which results in the observed Gaussian distribution of $G_s(r,t)$. No
such harmonic oscillations are possible in a lattice system and thus
the Gaussian is replaced by a $\delta-$function.

In view of these findings for $G_s(r,t)$ we consider the following simple
model for the dynamics of a tagged particle (similar ideas have been
explored in Refs.~\cite{barkai03,berthier05,jung05,heuer08}). 
At early times the
particle is locally trapped due to the presence of its neighbors. Hence
it makes harmonic oscillations with a typical amplitude $\ell$ and which
corresponds therefore to a Gaussian distribution of the displacement: $f_{\rm
vib}(r)=(2\pi \ell^2)^{-3/2} \exp(-r^2/2\ell^2)$. Thus if we start the
clock at $t=0$, at time $t_1$ the particle will have a displacement
that is drawn from the distribution $f_{\rm vib}$. At this time $t_1$
the particle makes a jump of length $\Delta$ and we assume that the
distribution of $\Delta$ is given by a Gaussian distribution of width $d$:
$f_{\rm jump}(\Delta)= (2\pi d^2)^{-3/2}\exp(-\Delta^2/2d^2)$. After
this jump the particle will again vibrate harmonically around its new
position and thus show a displacement that is distributed according to
$f_{\rm vib}(r)$. After a time $t_2$ it will make a new random hop by a
distance that is given by the distribution $f_{\rm jump}(\Delta)$. This
alternation of vibrations and jumps gives us the relaxation dynamics of
the particle. A model in which a particle makes an
alternation between sitting/waiting and hopping is called continuous
time random walk (CTRW)~\cite{montroll65} and in the past it has been
used to describe very different physical phenomena (hopping of atoms in
semiconductors, flight of birds, etc.)~\cite{klages08}.

Now, for any CTRW model, the distribution of the waiting times
$t_1$ can be calculated from the one of $t_2$~\cite{jung05}.
However, for glass-forming liquids, these distributions of
waiting times are very complex in nature, having significantly
non-exponential shapes.
It can be shown that, for such broad non-exponential 
distributions, $\langle{t_1}\rangle\ge\langle{t_2}\rangle$, where
$\langle{t_1}\rangle$ and $\langle{t_2}\rangle$ are the first moments
of the two distributions~\cite{jung06}. 
In the absence of any definite information
regarding the shape of the distribution functions and also for simplifying
calculations, we assume that the distributions $\phi_k$ for $t_1$ and
$t_2$ are independent from each other and describe both of them with a
single exponential, i.e. $\phi_k(t_k)=\tau_k^{-1} \exp(-t_k/\tau_k)$,
with $k=1,2$.  From such a dynamical model one can readily calculate
the self part of the van Hove function and one finds~\cite{montroll65}:

\begin{equation}
G_s(r,t)= \sum_{n=0}^\infty p(n,t) f(n,r) \quad .
\label{eq3}
\end{equation}

\noindent
Here $p(n,t)$ is the probability that the particle makes, within the time
$t$, exactly $n$ jumps and $f(n,r)$ is the probability that it makes in
$n$ jumps a displacement $r$. For the hopping dynamics of our CTRW model it
is not difficult to carry out the sum in Eq.~(\ref{eq3}). If one goes
into the Fourier-Laplace domain one finds

\begin{equation}
G_s(q,s)= f_{\rm vib}(q) \Phi_1(s) + f(q) f_{\rm vib}(q) 
\frac{\phi_1(s) \Phi_2(s)}{1-\phi_2(s) f(q)} \quad ,
\label{eq4}
\end{equation}

\noindent
where $\Phi_k := (1- \phi_k(s))/s$ and $f(q):= f_{\rm vib}(q) f_{\rm
jump}(q)$. (We mention that Eq.~(\ref{eq4}) is valid for any choice
of distribution ($f_{\rm vib}, f_{\rm jump}, \phi_1, \phi_2$), i.e. is
not restricted to the exponential and Gaussian distributions considered
here.)  Using the four free parameters $\ell, d, \tau_1$, and $\tau_2$
as fit parameters, we have used the expression~(\ref{eq4}) to fit the
data shown in Fig.~\ref{fig7_vanhoveall}. Note that i) a given choice
of fit parameter must allow to fit the data for {\it all} the times
considered and ii) that the inverse-Fourier transforms have to be done
numerically. The resulting fits are included in the figure as well
(solid lines) and we see that the agreement is indeed very good. Hence
we can conclude that, despite the simplicity of the model, it is able
to capture some of the characteristic of the relaxation dynamics of a
tagged particle in a glass-forming system. It is also of interest to
mention that the values for $\ell$ and $d$ that result from the fits are
very reasonable.  To a good approximation we find $d\approx 2 \ell$ and
($d/\sigma, \ell/\sigma$) is (0.1, 0.051) for colloids, (0.15, 0.06)
for grains, and (0.35, 0.15) for the Lennard-Jones system. Moreover, our
results for $d^2$ agree well with the height of the plateau measured in
the mean-squared displacements for all systems. Hence we can conclude
that the parameters of the model have values that are physically very
reasonable.

Finally we discuss the origin of the exponential tail in the distribution
function $G_s(r,t)$. For simplicity we consider the special case that
$\ell=0$, i.e. no vibrational motion occurs, and $\tau_1=\tau_2$. For
this choice one obtains the analytic result

\begin{equation}
G_s(r,t)= G_0 + \frac{4\pi e^{-t/\tau_1}}{r}
\int_0^\infty dq[e^{t f(q)/\tau_1} -1 ] q \sin(qr) \quad ,
\label{eq5}
\end{equation}

\noindent
where $G_0(r,t) \equiv \delta(r) \Phi_1(t)$. By expanding the exponential
in Eq.~(\ref{eq5}) in a power series, integrating each term and converting 
the resulting sum into an integral one obtains

\begin{equation}
G_s(r,t) = G_0(r,t) + \frac{\pi e^{-t/\tau_1}}{4 d^3}
\int_1^\infty dn \frac{e^{-f(n)}}{n^2} \quad ,
\label{eq6}
\end{equation}

\noindent
with $f(n) := n \ln n -n \ln (t/\tau_1) -n +r^2/(8d^2 n)$. For large
$r$, Eq.~(\ref{eq6}) can be evaluated using a saddle point approximation
and then gives

\begin{equation}
G_s(r,t) \sim \frac{(\pi Y)^{3/2} e^{-t/\tau_1}}{(rd)^{3/2} \sqrt{1+Y^2}}
e^{-r[Y-1/Y]/2d} \quad ,
\label{eq7}
\end{equation}

\noindent
where $Y$ is defined via the equation $Y^2 \exp Y^2 = r^2/(2dt/\tau_1)^2$,
i.e. $Y^2 \sim 2 \ln (r^2/(2dt/\tau_1))$ if $r$ is large. Thus we find
that $G_s(r,t)$ has for large $r$ indeed an exponential tail (with
logarithmic corrections), as found in the simulation and experimental
data (see Fig.~\ref{fig7_vanhoveall}). We emphasize that the asymptotic
result given by Eq.~(\ref{eq7}) can be obtained for any distribution
of $f_{\rm vib}(r), f_{\rm jump}, \phi_{1}, \phi_2(t_2)$, i.e. it is a
quite general result. The exponential tail is thus a direct consequence of
the fact that the timescales between the jumps is a distributed quantity
and so it directly follows from the dynamically heterogeneous behavior
observed in glass-formers.

\section{Conclusions}

In this paper we have discussed the relaxation dynamics of a lattice gas
model in which the only non-trivial features are the rules governing the
motion of the particles. Despite its simplicity, this model has a dynamics
that is similar to the one of real glass-forming systems in that it shows
a strong slowing down with increasing density and a stretching of time
correlation functions. By tracking the trajectory of a tagged particle,
we found that its dynamics is very heterogeneous in space and time, in
agreement with the dynamics of real glass-formers, for which, however,
the existence of dynamical heterogeneity can usually be inferred from
experimental data only in an indirect way. We find that the regions in
which the particles move form, at early stages, isolated blobs, which,
with increasing $t$, grow and start to percolate. The resulting spanning
structure allows some of the particles to make relatively quickly large
displacements, a behavior that gives rise to an exponential tail in
the self part of the van Hove function. At very large times, i.e. much
larger than the $\alpha-$relaxation time, the cluster of active sites
fills the whole space and thus the system has completely relaxed.

Although not discussed in the present manuscript, it is possible
to predict to some extent at which lattice sites the blobs and the
percolating cluster will occur. In Ref.~\cite{chaudhuri08} it was
shown that these active sites are strongly correlated with lattice
sites that have, at time $t=0$, a relatively low local density of
particles. Hence we can conclude that, for this model, the initial
configuration of the particles allows to predict the details of the
relaxation dynamics of the system at short and intermediate times.

Finally we have shown that the exponential tail of the van Hove function
can be rationalized by a simple model based on a continuous time
random walk and that this model can be used successfully to describe
also the space and time dependence of $G_s(r,t)$ for off-lattice
models. Since at the moment this model is purely phenomenological,
it will be interesting to see whether it is possible to derive
it from a more microscopic many-body theory such as mode-coupling
theory~\cite{gotze92,gotze99,das04}. Last but not least it is also important
to investigate better to what extent the percolating cluster that we find
to be important for the relaxation dynamics of the lattice gas discussed
here, is present also in off-lattice system. Thus it is evident that
despite the present results, there are still many open and interesting
questions regarding the relaxation dynamics of glass-forming systems.

\begin{theacknowledgments}
We thank O. Dauchot, G. Marty, and E. Weeks for providing
their data, G. Biroli, J.-P. Bouchaud, P. Mayer, and D. Reichman for useful
discussions. Financial support from CEFIPRA Project 3004-1, and ANR
Grant TSANET are acknowledged.
\end{theacknowledgments}

\end{document}